\documentclass[review]{elsarticle}
\usepackage{color}
\usepackage{amssymb}
\usepackage{lineno,hyperref}
\modulolinenumbers[5]

\journal{Journal of \LaTeX\ Templates}

\newcommand{\jpsi}{\mathrm{J/}\psi}

\newcommand{\Wgp}{W_{\gamma\mathrm{p}}}

\begin{document}

\begin{frontmatter}

\title{Energy dependence of dissociative $\jpsi$ photoproduction as a signature of gluon saturation at the LHC}

\author[CVUT]{J. Cepila}
\author[CVUT]{J. G. Contreras}
\author[KU]{J. D. Tapia Takaki}
\address[CVUT]{Faculty of Nuclear Sciences and Physical Engineering,
Czech Technical University in Prague, Czech Republic}
\address[KU]{Department of Physics and Astronomy, The University of Kansas, Lawrence, KS, USA}

\begin{abstract}
We have developed a model in which the quantum fluctuations of the proton structure are characterised by hot spots, whose number  grows with decreasing Bjorken-$x$. Our model reproduces  the $F_2(x,Q^2)$ data from HERA at the relevant scale, as well as the exclusive and dissociative $\jpsi$ photoproduction data from H1 and ALICE. Our model predicts that for  $\Wgp \approx 500$ GeV, the dissociative $\jpsi$ cross section reaches a maximum and then decreases steeply with energy, which is in qualitatively good agreement to a recent observation that the dissociative $\jpsi$ background in the exclusive $\jpsi$ sample measured in photoproduction by ALICE decreases as energy increases. Our prediction provides a clear signature for gluon saturation at LHC energies. 
\end{abstract}

\begin{keyword}
Gluon saturation, vector meson photoproduction, LHC
\end{keyword}

\end{frontmatter}


\section{Introduction}

Perturbative QCD predicts that the gluon density in hadrons grows with  energy, or equivalently with decreasing Bjorken-$x$, up to a point where non-linear effects set in to tame this growth, a phenomenon known as gluon saturation; see for example~\cite{Gelis:2010nm} and references therein.

Inclusive data on the proton structure function $F_2(x,Q^2)$ -- where $Q^2$ is the virtuality of the photon -- from deeply inelastic scattering at HERA~\cite{Abramowicz:2015mha} show that indeed the gluon density grows steeply at small $x$.  At the same time the analysis of these data is inconclusive regarding the question of gluon saturation, because the data can be equally well described without, e.g.~\cite{Abramowicz:2015mha}, or with, e.g.~\cite{Albacete:2010sy}, the inclusion of saturation effects. 

\begin{figure}[!t]
\centering 
\includegraphics[width=0.48\textwidth]{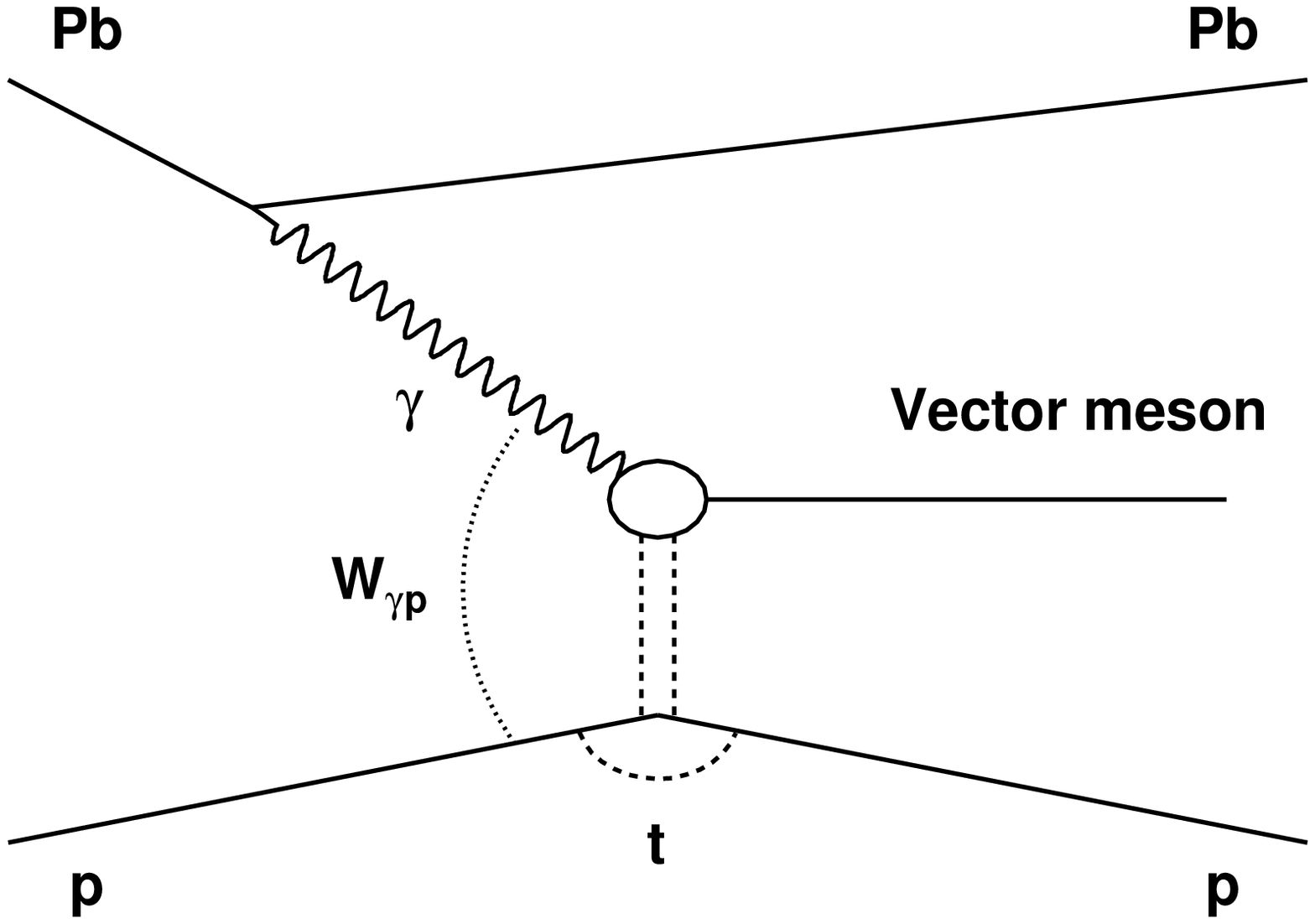}
\includegraphics[width=0.48\textwidth]{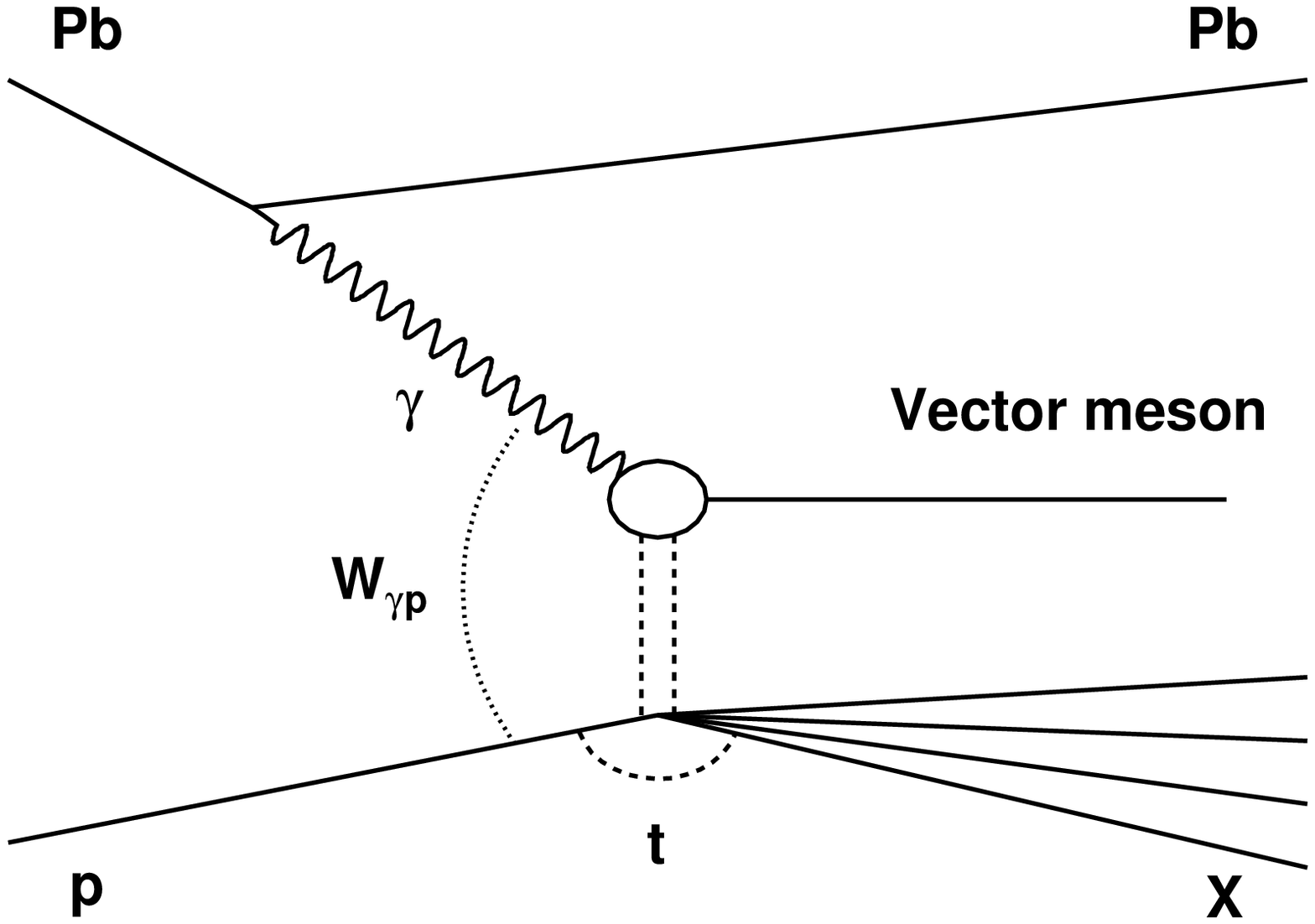}
\caption{
\label{fig:VMP} 
Diagrams for exclusive (left) and dissociative (right) vector meson photo-production.  The source of photons is a lead nucleus as  in p-Pb collisions at the LHC. For the case of HERA, the source of photons was either an electron or a positron.
}
\end{figure}

$F_2(x,Q^2)$  data are sensitive to longitudinal degrees of freedom. Exclusive vector meson photoproduction, see left panel of Fig~\ref{fig:VMP}, is also sensitive to the distribution of gluons in the impact parameter plane, through the  dependence of the cross section on $t$, the square of the momentum transfer at  the target vertex. This process has been extensively studied at HERA~\cite{Ivanov:2004ax}. For a recent review see~\cite{Newman:2013ada}. For the case of $\jpsi$ photoproduction, HERA measurements cover a range on $\Wgp$, the centre-of-mass energy of the photon-proton system, from 20 to 300 GeV, which corresponds to $10^{-4}\lesssim x \lesssim 0.02$, where  $x=M_{\jpsi}^2/\Wgp^2$, with $M_{\jpsi}$ as the mass of the $\jpsi$.  ALICE has also published measurements of exclusive $\jpsi$ photoproduction in p-Pb collisions at the LHC at low energies,   20 GeV $<\Wgp<$ 40 GeV, and at high energies $\langle \Wgp\rangle=700$ GeV~\cite{TheALICE:2014dwa}. This last measurement corresponds to $x=2\cdot 10^{-5}$. As for the $F_2(x,Q^2)$   data, the exclusive $\jpsi$ photoproduction data can  be described with~\cite{Armesto:2014sma} or without~\cite{Jones:2013pga}  saturation effects.

In a Good-Walker formalism~\cite{Good:1960ba} exclusive diffractive processes are sensitive to the average over the different configurations of the target, while dissociative processes, where the target get excited and produces many particles, measure the variance over the configurations~\cite{Miettinen:1978jb}. 
Dissociative vector meson photoproduction,  see right panel of Fig~\ref{fig:VMP}, has also been measured at HERA. Recently, H1 measured simultaneously the exclusive and dissociative photoproduction of $\jpsi$ and reported both, the $\Wgp$ and the $t$ dependence of these cross sections~\cite{Alexa:2013xxa}. 

The $t$ dependence of dissociative vector meson photoproduction at a fixed value of $\Wgp$ has been studied  in a model where the transverse distribution of gluons in the proton is described as the sum of three independent gaussian distributions~\cite{Mantysaari:2016ykx,Mantysaari:2016jaz}. In that analysis it was shown that a good description of data is achieved when the geometrical position of these gaussian distributions fluctuate from event to event.

 In this Letter we study the energy dependence of dissociative $\jpsi$ photoproduction off protons, showing that it provides a  clear signature of gluon saturation at the LHC. Our calculation reproduces correctly the rise with $\Wgp$ of this cross section as measured at HERA~\cite{Alexa:2013xxa}, and predicts that it reaches a maximum at $\Wgp\approx 500$ GeV,   followed by a steep decrease at higher energies. In our model,  as energy increases the different configurations of the proton resemble more and more each other, so the variance over configurations decreases. The LHC can explore  a wide range of $\Wgp$, from a few tens of GeV to more than a TeV, making it an ideal place to look for this clear signature of gluon saturation.

The rest of this Letter is organised as follows. In the next section we present all the formulae used in this work. Section~\ref{sec:Results} explains how the parameters have been fixed and compares the predictions of our model to experimental data. In Section~\ref{sec:Discussion} we comment on the model, present our main conclusion and remark on other potentially interesting measurements that can be carried out to test our model. Section~\ref{sec:Summary} gives a brief summary.
 
\section{The amplitude for diffractive photoproduction of vector mesons
\label{sec:amplitude}}

Diagrams for the exclusive and dissociative production of vector mesons are shown in Fig.~\ref{fig:VMP} where the photon source  is a lead ion, which corresponds to the ideal case to study vector meson photoproduction off the proton at the LHC~\cite{Contreras:2015dqa}. At HERA the source was either a positron or an electron.
In the dipole colour model~\cite{Mueller:1989st,Nikolaev:1990ja}, the interaction of the photon with the target proceeds in steps: first the photon fluctuates into a quark-antiquark pair, then this pair interacts with the target with the colour dipole cross section $\sigma_{\rm dip}$ and finally the dipole forms a vector meson. In this approach the amplitude can be written as follows (see for example~\cite{Kowalski:2006hc}),

\begin{equation}
A(x,Q^2,\vec{\Delta})_{T,L} = i\int d\vec{r}\int^1_0\frac{dz}{4\pi}
(\Psi^*\Psi_{\rm V})_{T,L} \int d\vec{b}\;
e^{-i(\vec{b}-(1-z)\vec{r})\cdot\vec{\Delta}}\frac{d\sigma_{\rm dip}}{d\vec{b}},
\label{eq:Amplitude}
\end{equation}
where $-t = \vec{\Delta}^2$, $\Psi_{\rm V}$ is the wave function of the vector meson, $\Psi$ is the wave function of a virtual photon fluctuating into a quark-antiquark colour dipole, $\vec{r}$ is the transverse distance between the quark and the antiquark, $z$ is the fraction of the longitudinal momentum of the dipole carried by the quark and $\vec{b}$ is the
impact parameter. The subindices $T$ and $L$ refer to the contribution of the transversal and longitudinal degrees of freedom of the virtual photon, respectively. 

The physics of the interaction is encoded in the  dipole-target cross section, which is related, via the optical theorem, to the imaginary part of the forward dipole-target amplitude $N(x,\vec{r},\vec{b})$ :

\begin{equation}
\frac{d\sigma_{\rm dip}}{d\vec{b}} = 2 N(x,\vec{r},\vec{b}).
\label{eq:b-depSigDip}
\end{equation}

The main goal of our analysis is to study the effect of fluctuations  in the configuration of the proton using   the energy dependence of the dissociative cross section. In order to isolate these two contributions  we use a factorised form for the dipole amplitude:
\begin{equation}
N(x,r,b) = \sigma_0N(x,r)T(\vec{b}).
\label{eq:Nxrb}
\end{equation}
Here, $r$ and $b$ are the magnitudes of $\vec{r}$ and $\vec{b}$, respectively; $\sigma_0$ is a constant that  provides the normalisation and $T(\vec{b})$ describes the proton profile in the impact parameter plane. Such a  factorised model has been successfully used in the past; e.g. in~\cite{Marquet:2007nf}. There are several possibilities to chose a model for $N(x,r)$; e.g., ~\cite{Iancu:2003ge,Albacete:2010sy}. For simplicity, we chose the form of $N(x,r)$ given by the model of Golec-Biernat and Wusthoff~\cite{GolecBiernat:1998js}:
   
\begin{equation}
 N(x,r) = \left( 1-e^{-r^2Q^2_s(x)/4}\right),
\end{equation}
with the saturation scale given by
\begin{equation}
Q^2_s(x) = Q^2_0(x_0/x)^\lambda,
\end{equation}
where $\lambda$, $x_0$ and $Q^2_0$ are parameters. The latter is normally set to 1 GeV$^2$ as done in this study.

The proton is a quantum object, so its structure fluctuates from interaction to interaction. In our model all fluctuations are included in the proton profile $T(\vec{b})$. Following~\cite{Mantysaari:2016ykx,Mantysaari:2016jaz}  we define the proton profile as the sum of $N_{hs}$ regions of high gluon density, so called hot spots,  each of them having a gaussian distribution:
\begin{equation}
T(\vec{b}) = \frac{1}{N_{hs}}\sum^{N_{hs}}_{i=1}T_{hs}(\vec{b}-\vec{b_i}),
\end{equation}
with
\begin{equation}
T_{hs}(\vec{b}-\vec{b_i}) = \frac{1}{2\pi B_{hs}}e^{-\frac{(\vec{b}-\vec{b_i})^2}{2B_{hs}}},
\end{equation}
where each $\vec{b_i}$ is obtained from a 2-dimensional gaussian distribution centred at $(0,0)$ and having a width $B_p$. The parameters $2B_p$ and $2B_{hs}$ can be interpreted as  the average of the squared transverse
radius of the proton or the hot spots, respectively. Such a profile has already been used before in a study of nuclear targets~\cite{Caldwell:2010zza}.

The new ingredient in our model is that we introduce an indirect energy dependence of the proton profile $T(\vec{b})$ by making the number of hot spots grow with decreasing $x$. That is,
$N_{hs}=N_{hs}(x)$, where the exact definition used in our model is given below
in Eq.~(\ref{eq:nhs}).
This implements the hypothesis that, at a given fixed scale, as energy increases, the number of gluons available for the interaction increases.

Putting together all ingredients, and taking advantage that some of the integrals can be performed analytically, the amplitude can be written as 
\begin{equation}
A(x,Q^2,\vec{\Delta})_{T,L} = i A_b(\vec{\Delta})A_r(x,Q^2,\vec{\Delta})_{T,L},
\label{eq:factAmplitude}
\end{equation}
where,
\begin{eqnarray}
A_b &\equiv&  \int d\vec{b} e^{-i\vec{b}\cdot\vec{\Delta}}T(\vec{b}) \\
   &=& e^{-B_{hs}\Delta^2/2}\frac{1}{N_{hs}}\sum^{N_{hs}}_{i=1} e^{-i\vec{b_i}\cdot\vec{\Delta}}.
\end{eqnarray}
and, after integration over the angular variable,
\begin{equation}
A_r(x,Q^2,\Delta)_{T,L} \equiv  \int dr r N(x,r)  A_z(r,\Delta)_{T,L},
\label{eq:Ar}
\end{equation}
with 
\begin{equation}
A_z(r,\Delta)_{T,L} \equiv  \int dz (\Psi^*\Psi_{\rm V})_{T,L}J_0(r(1-z)\Delta).
\end{equation}

Using this amplitude, the cross sections are
\begin{equation}
\left.
\frac{d\sigma(\gamma p\rightarrow \jpsi p)}{dt}
\right|_{T,L} =
 \frac{(R^{T,L}_g)^2}{16\pi}
 \left| 
\left< A(x,Q^2,\vec{\Delta})_{T,L}\right>
\right|^2,
\label{eq:ExclXS}
\end{equation}
for the exclusive process, and
\begin{equation}
\left.
\frac{d\sigma(\gamma p\rightarrow \jpsi Y)}{dt} 
\right|_{T,L}= \frac{(R^{T,L}_g)^2}{16\pi}
\left(\left<\left| 
A(x,Q^2,\vec{\Delta})_{T,L}
\right|^2\right>
-
\left| 
\left< A(x,Q^2,\vec{\Delta})_{T,L}
\right>\right|^2\right),
\label{eq:DissXS}
\end{equation}
 for dissociative production,
where $R^{T,L}_g$ is the so-called skewedness correction~\cite{Shuvaev:1999ce} given by
\begin{equation}
R^{T,L}_g(\lambda^{T,L}_g) = \frac{2^{2\lambda^{T,L}_g+3}}{\sqrt{\pi}}\frac{\Gamma(\lambda^{T,L}_g+5/2)}{\Gamma(\lambda^{T,L}_g+4)},
\label{eq:Rg}
\end{equation}
with $\lambda^{T,L}_g$ given by 
\begin{equation}
\lambda^{T,L}_g\equiv \frac{\partial\ln(A_{T,L})}{\partial\ln(1/x)}.
\label{eq:lambda}
\end{equation}
\noindent The total cross section is the sum of the  $T$ and $L$ contributions.

We also compute  $F_2(x,Q^2)$, for which we use
\begin{equation}
F_2(x,Q^2)=\frac{Q^2}{4\pi^2\alpha_{em}}\Bigg(\sigma^{\gamma^\ast p}_{T}(x,Q^2)+\sigma^{\gamma^\ast p}_{L}(x,Q^2)\Bigg),
\label{eq:F2}
\end{equation}
where $\alpha_{em}$ is the electromagnetic coupling constant and
\begin{equation}
\sigma^{\gamma^\ast p}_{T,L}(x,Q^2)= \sigma_0 \int d \vec{r} \int^1_0 dz \big |\Psi_{T,L}(z,r,Q^2)\big |^2 N(r,\tilde{x}),
\label{eq:Sgp}
\end{equation}
where, using the approach of~\cite{GolecBiernat:1998js}, we use $\tilde{x} = x( 1+(4 m^2_f)/Q^2)$ with $m_f$ an effective mass for light quarks set to 140 MeV. 
\section{Results: comparison to data and predictions
\label{sec:Results}}

For $\Psi_{T,L}$ we use the definitions and parameter values of~\cite{GolecBiernat:1998js} and for the wave function of the vector meson, $\Psi_{\rm V}$,  we use the boosted gaussian model ~\cite{Nemchik:1994fp,Nemchik:1996cw}, with the numerical values of the parameters  as in~\cite{Kowalski:2006hc}. 

\begin{figure}[!t]
\centering 
\includegraphics[width=0.48\textwidth]{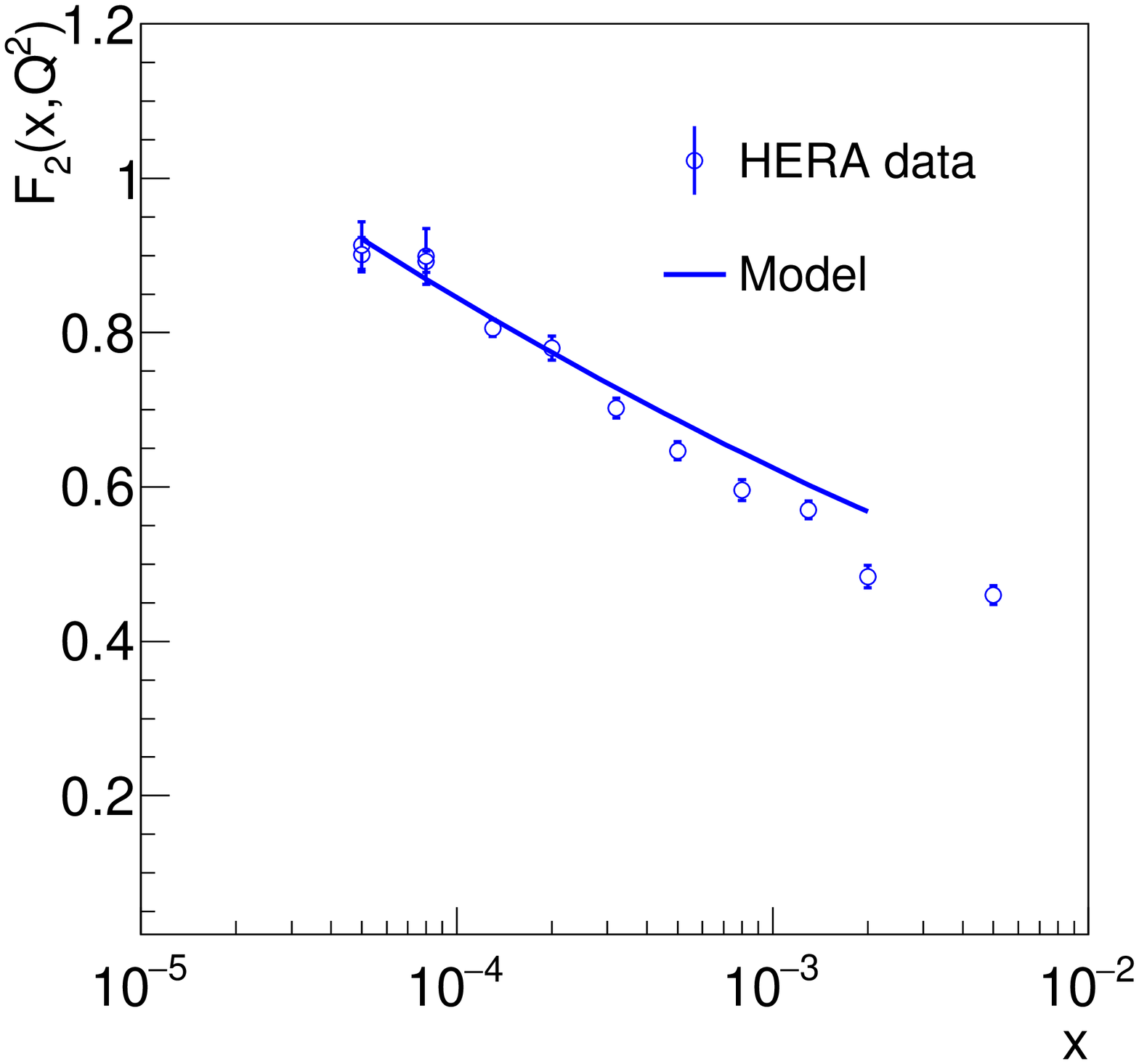}
\includegraphics[width=0.48\textwidth]{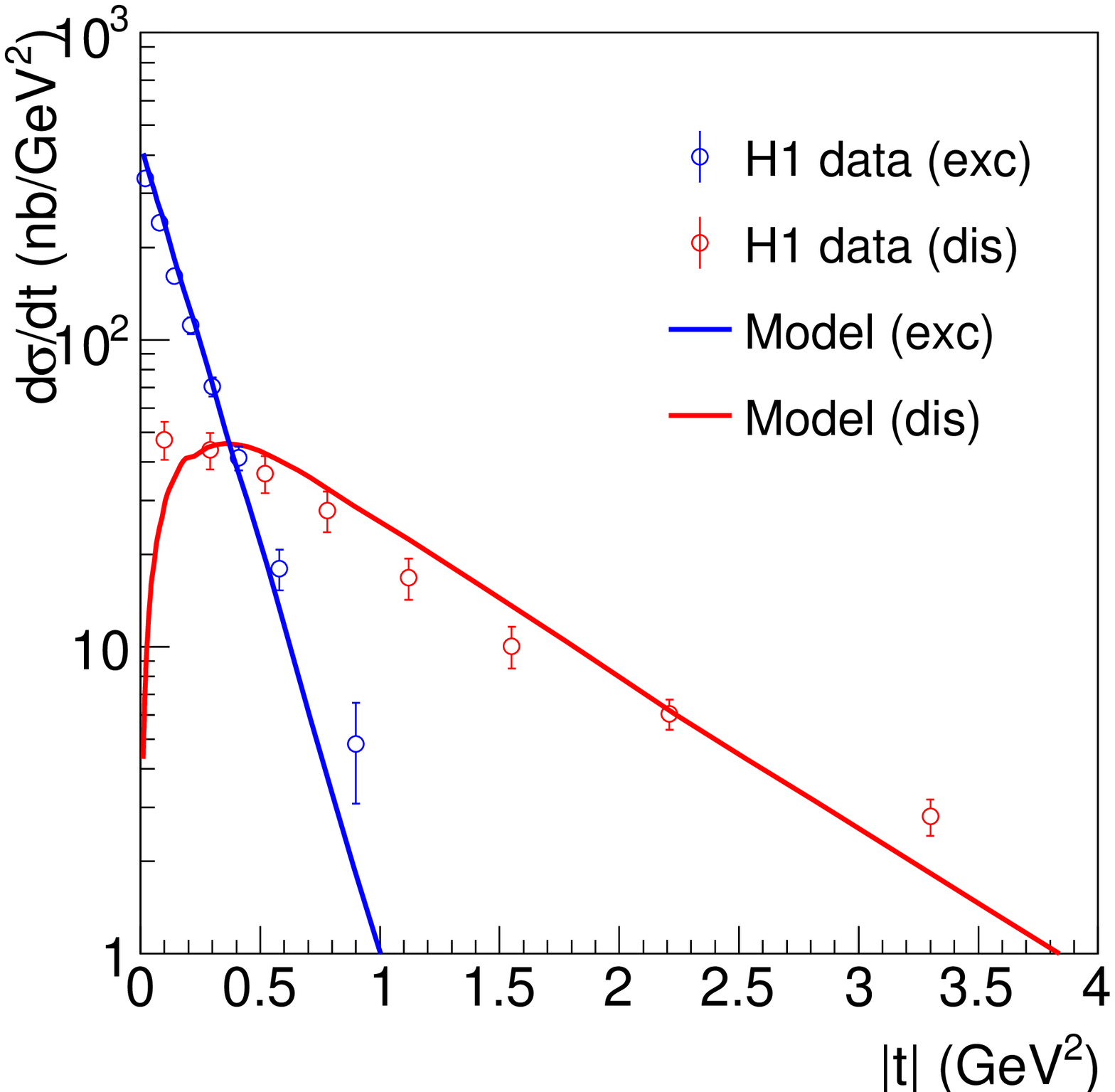}
\caption{
\label{fig:Comp} 
Comparison of the model (solid lines) to data (open bullets) on (left) the structure function of the proton $F_2(x,Q^2)$ at $Q^2=2.7$ GeV$^2$ as measured by  H1 and Zeus~\cite{Aaron:2009aa}  and (right) the $|t|$ distribution of exclusive (blue) and dissociative (red) photoproduction of $\jpsi$ as measured by H1~\cite{Alexa:2013xxa} at $\langle\Wgp\rangle=78$ GeV. 
}
\end{figure}

\begin{figure}[!t]
\centering 
\includegraphics[width=0.48\textwidth]{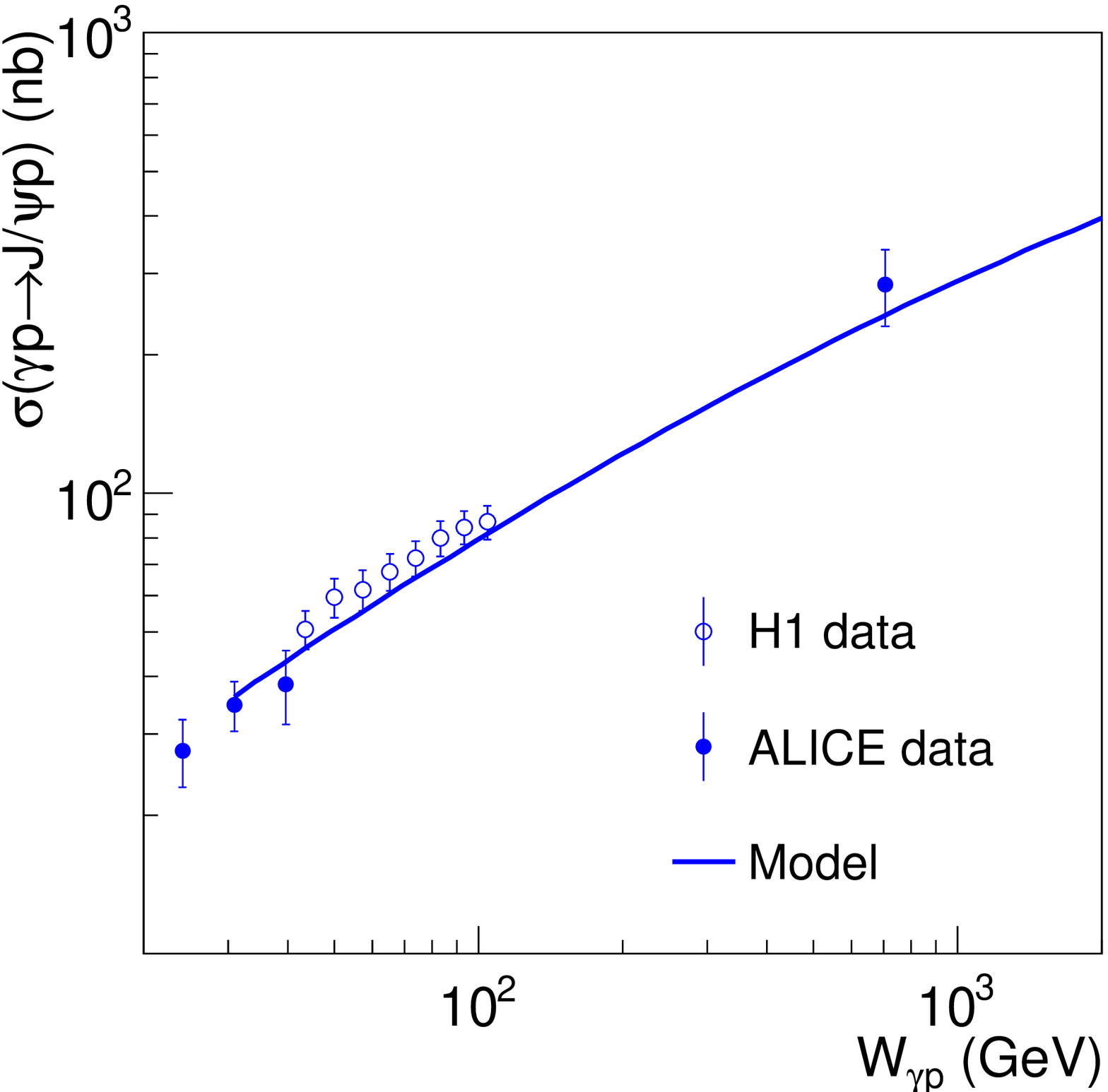}
\includegraphics[width=0.48\textwidth]{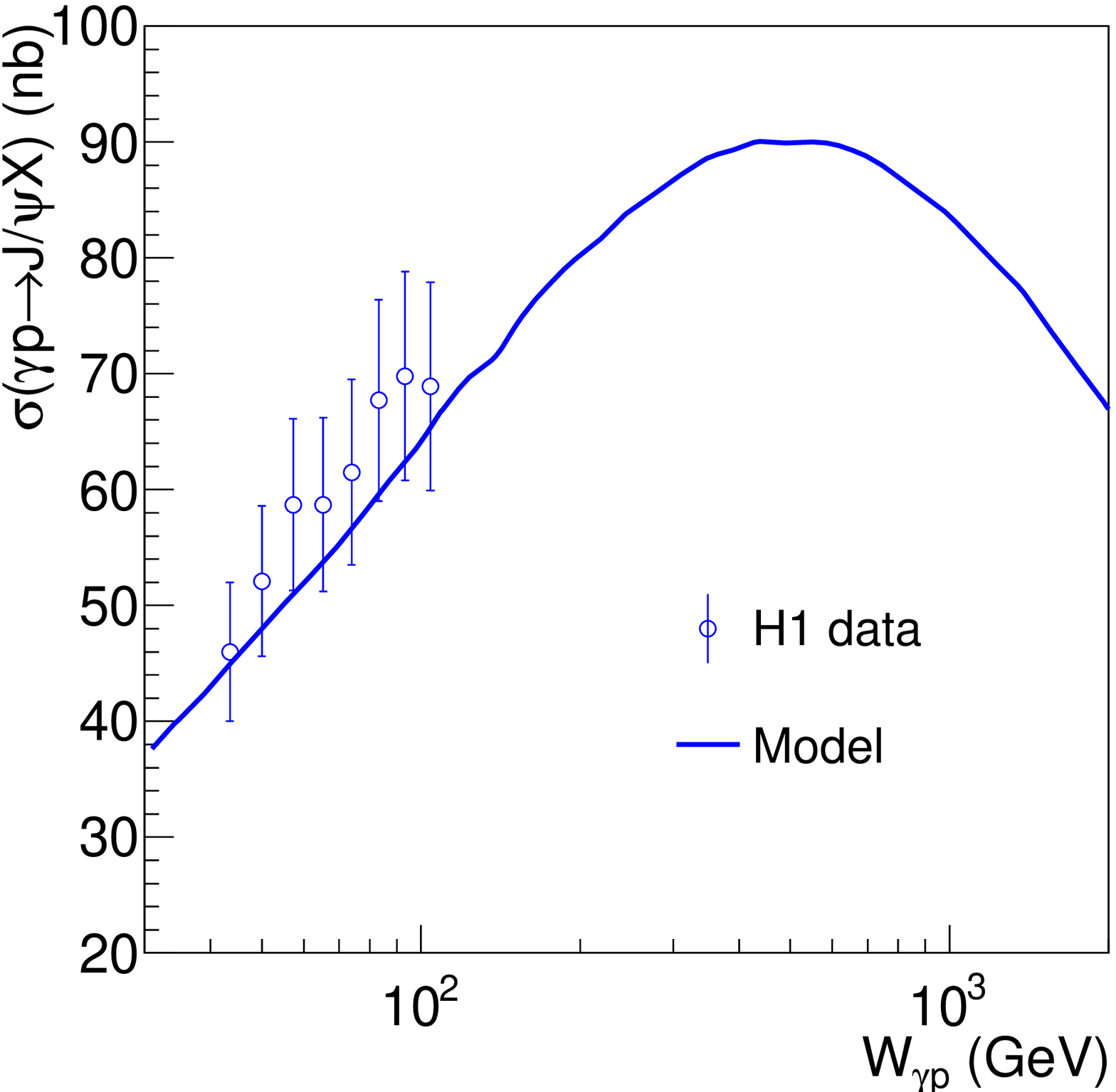}
\caption{
\label{fig:Pred} 
Comparison of the model (solid lines) to data  on the $\Wgp$ dependence of the cross section for exclusive (left) and dissociative (right) photoproduction of $\jpsi$ as measured by H1~\cite{Alexa:2013xxa} and ALICE~\cite{TheALICE:2014dwa} (open and solid bullets, respectively). 
}
\end{figure}

The parameters of the model were fixed as follows. The value of $\lambda$ is constrained by the energy dependence of  exclusive $\jpsi$  photoproduction to $\lambda=0.21$. The parameter $B_p$ is constrained by the $t$ distribution of the same data and set to $B_p=4.7$ GeV$^{-2}$. We define $\sigma_0\equiv4\pi B_p$. The $x_0$ parameter is strongly correlated to $\sigma_0$, so once $B_p$ is fixed, the normalisation of the data yields $x_0=0.0002$. The parameter $B_{hs}$ is constrained by the $t$ dependence of the dissociative process and it is set to $B_{hs}=0.8$  GeV$^{-2}$. Finally, as we relate the number of hot spots with the number of gluons available for the interaction, we follow standard functional forms for the gluon distribution as used in PDF fits; e.g.~\cite{Martin:2009iq}, and parametrise
\begin{equation}
N_{hs}(x) = p_0x^{p_1}(1+p_2\sqrt{x}),
\label{eq:nhs}
\end{equation}
where we set $p_0=0.011$, $p_1=-0.58$ and $p_2=250$ in order to reproduce the energy dependence of H1 measurements of dissociative $\jpsi$ photoproduction. The parameters $p_0$ and $p_1$ are strongly correlated. For the quoted value of $p_0$, a similar description of data is found when varying the value of $p_1$ by $\pm$5\% and of $p_2$ by $\pm$15\%. Note that there is a trade off between the normalisation of the data in the right
panel of Fig.~\ref{fig:Pred} and that of the dissociative data in the right panel of Fig.~\ref{fig:Comp}.
Improving the normalisation in the first case, would worsen unacceptably the
description of the $|t|$ distribution of the dissociative cross section. The chosen
parameters are those that yield  a correct simultaneous description of the
data in both figures. With these parameters a $\chi^2/d.o.f\approx 1$ is obtained for the comparison of the model predictions and the data from H1 on the energy dependence of the exclusive and dissociative cross sections.

The comparison of the model with HERA data for $F_2(x,Q^2)$ at $Q^2=2.7$ GeV$^2$ as measured by  H1 and Zeus~\cite{Aaron:2009aa} is shown in the left panel of Fig.~\ref{fig:Comp}. 
This scale is chosen since it is similar to that of $\jpsi$ photoproduction, which is often set 
 to $(M_{\jpsi}/2)^2=2.4$ GeV$^2$~\cite{Ryskin:1992ui}.
 Although the model is relatively simple and has been developed to describe vector meson photoproduction, it describes well the $F_2(x,Q^2)$ data. 
The comparison  with H1 data~\cite{Alexa:2013xxa} on the $t$ dependence of $\jpsi$ photoproduction is shown in the right panel of the same figure for the data at $\langle\Wgp\rangle=78$ GeV. Both the exclusive and the dissociative processes are reasonably well described. 
The left panel of Fig.~\ref{fig:Pred} shows the energy dependence of the model and compares it to data from H1~\cite{Alexa:2013xxa} and from ALICE p-Pb collisions at a 5.02 TeV centre-of-mass energy~\cite{TheALICE:2014dwa}. Again, the model describes quite well the data. 

Our main result is shown in the right panel of  Fig.~\ref{fig:Pred}. We show that the model describes correctly the energy evolution of the dissociative cross section for the available data. We also show that the model {\it  predicts} that the cross section will reach a maximum at $\Wgp\approx 500$ GeV and it will then turn around and decrease for higher energies.
This decrease of the dissociative $\jpsi$ photoproduction cross section is significantly large and can be observed in the $\Wgp$ energy range accessible at the LHC.

\section{Discussion
\label{sec:Discussion}}

Several comments are in order. The chosen numerical values of the parameters are quite reasonable: $B_p$ is similar to that measured at HERA~\cite{Alexa:2013xxa} and $\lambda$ is similar to the value found at HERA for a scale $Q^2\approx2-3$ GeV$^2$~\cite{Aaron:2009bp}. Regarding the value of $B_{hs}$, it corresponds to a hot spot radius of 0.35 fm, quite close to the values around 0.3 fm discussed in~\cite{Braun:1992jp,Kopeliovich:2000pc,Kovner:2002xa}.

In our model the energy dependence  shown in the right panel of Fig.~\ref{fig:Pred} has a geometrical origin reminiscent of percolation. At some point the  number of hot spots is so large that they overlap. When the overlap is large enough, different configurations look the same  and the variance (see Eq.~(\ref{eq:DissXS})) decreases. A percolation approach has been explored for the fusion of strings yielding a quark gluon plasma~\cite{Armesto:1996kt}, where interestingly the radius of the strings is also found to be around 0.3 fm. For the parameters that we have employed in this study, the maximum at $\Wgp\approx$ 500 GeV is reached when some 10 hot spots are present, indicating a sizeable overlap of hot spots.

In conclusion, the main result is that our model predicts that 
the dissociative photoproduction cross section decreases at  high energies, which are reachable at the LHC. Furthermore, experimental data from ($i$) ALICE~\cite{TheALICE:2014dwa} and ($ii$) H1\cite{Alexa:2013xxa}  suggests that this is indeed so, which we take as a further support of the chosen parameters of our model:

($i$) ALICE has measured the cross section for exclusive $\jpsi$ photoproduction in
p-Pb collisions~\cite{TheALICE:2014dwa}. The main background to this process is dissociative $\jpsi$
photoproduction. For the conditions existing in ALICE, $|t|$ is related to the square
of the transverse momentum of the $\jpsi$ ($p_T$). The dissociative
contribution populates the large $|t|$ region (see right panel of Fig.~\ref{fig:Comp}) so in ALICE
it populates the large $p_T$ region.
This contribution is clearly present in the upper panel of Fig. 2 in Ref.~\cite{TheALICE:2014dwa} which correspond to $\Wgp$  ranging from 20 to 40 GeV, but it is almost absent in the lower panel of Fig. 2 in Ref.~\cite{TheALICE:2014dwa} depicting the high energy data sample at $\langle\Wgp\rangle$  = 700 GeV.  Qualitatively, the behaviour of these ALICE data is similar to that of our predictions shown in the left panel of Fig.~\ref{fig:Pred}. We take this observation as an indication that the prediction of our model is qualitatively correct.
Indeed, the behaviour of the background from dissociative $\jpsi$ photoproduction observed by ALICE seems to indicate that the turn around may be even at smaller energies. 

($ii$) Inspecting H1 data, specifically Table 2 of~\cite{Alexa:2013xxa}, one can observe  that the reported cross sections at the highest values of $\Wgp$ are the same for the last three measurements, albeit with somewhat large errors. This seems to suggest that the growth of the dissociative cross section with decreasing $x$ may already be slowing down at HERA energies.

In our model the shape of the $t$ dependence of the cross sections for the dissociative and exclusive processes does not vary with $x$.
 However, if the dissociative $\jpsi$ cross section disappears at relatively high energies as predicted in our model, it will simplify subtracting the dissociative background in the measurement of the $t$ distribution of exclusive vector meson photoproduction, particularly at large $t$ values. This is important since the appearance of diffractive dips in the $t$-distirbution at large $t$ has been proposed as a good observable for gluon saturation effects~\cite{Toll:2012mb,Armesto:2014sma}.

In addition, it will be interesting to study, experimentally and phenomenologically, the corresponding process with nuclear targets at the LHC. Both the $t$ and the $W_{\gamma{\rm Pb}}$ dependence may offer interesting insights  since  nuclear effects are expected to be significantly stronger than in $\gamma$-proton interactions~\cite{Albacete:2014fwa}:
At a given rapidity, coherent $\jpsi$ photoproduction in Pb-Pb reaches values
of $x$ almost a factor three larger than in p-Pb, because the energy of the
incoming proton beam is larger that that of the incoming Pb beam. This implies,
see Eq.~(\ref{eq:nhs}), that $N_{hs}(x)$ is smaller for the nucleons in Pb targets, but as
there are so many nucleons, the {\it total} number of hot-spots is greatly
increased.
Existing results from ALICE~\cite{Abelev:2012ba,Abbas:2013oua} and CMS~\cite{Khachatryan:2016qhq} on coherent $\jpsi$ photonuclear production are very promising. 

In principle, such a process can also be studied in proton-proton collisions at
the LHC. The advantages would be a higher centre-of-mass energy and a larger
available luminosity. The disadvantages would be a much smaller photon flux
from a proton with respect to that from a lead ion and the fact that it is not
known which proton act as a source. It could be argued that the dissociative
system can be used to tag the target proton, but in proton-proton collisions the
probability of extra exchange of soft gluons is important (see e.g.~\cite{Schafer:2007mm}) and these gluons
could produce the dissociation of the proton. As these are soft processes they
cannot be completely computed in pQCD and have to be modelled, adding extra
uncertainties to the extraction of photoproduction cross sections. Nonetheless,
first measurements of the exclusive contribution have already been performed
by LHCb and a model dependent extraction of the exclusive cross section has been
attempted~\cite{Aaij:2014iea}.

As a final remark, we note that in Fall 2016, the LHC Collaborations are collecting   proton-lead data at 8 TeV for the first time. These data are expected to test the predictions presented in this Letter. The unambiguous discovering of gluon saturation effects  at the LHC, would boost the proposal for future electron-proton and electron-ion colliders~\cite{Accardi:2012qut,AbelleiraFernandez:2012cc,Geesaman:2015fha}, where this phenomenon could be studied with much more precision and with many more observables than at the LHC.

\section{Summary and outlook
\label{sec:Summary}}

We have presented a model  for the dissociative $\jpsi$ photoproduction cross section  The model incorporates a fluctuating hot spot structure of the proton in the impact parameter plane. The new ingredient of the model is that the number of hot spots grows with decreasing $x$. The model describes correctly the behaviour of $F_2(x,Q^2)$ at the relevant scale, as well as  the $\Wgp$ and $t$ dependence of the exclusive and dissociative $\jpsi$ photoproduction cross section as measured by H1 and ALICE. The model predicts that the energy dependence of the dissociative process increases from low energies up to $\Wgp\approx500$ GeV and then decreases steeply. This behaviour happens within the energy range accessible at the LHC.

\section*{Acknowledgements}
This work was partially supported by grant LK11209 of  M\v{S}MT \v{C}R and
 by  grant 13-20841S of the Czech Science Foundation (GACR).

\section*{References}

\bibliography{ALICE,QCD,HERA,CMS,LHCothers}

\end{document}